\documentclass{kapproc} 
\usepackage{epsfig}

\normallatexbib

\begin{document}

\articletitle{Ratchet effects in Luttinger liquids}


\author{D. E. Feldman}

\affil{Department of Physics, Brown University, 182 Hope Street,
  Providence, RI 02912, USA;
  \\
  Landau Institute for Theoretical Physics, Chernogolovka, Moscow
  region 142432, Russia}

\email{Dima$\_$Feldman@holley.physics.brown.edu}

\author{S. Scheidl}

\affil{Institut f\"ur Theoretische Physik, Universit\"at zu K\"oln,
  Z\"ulpicher Strasse 77, 50937 K\"oln, Germany}

\email{sts@thp.uni-koeln.de}

\author{V. M. Vinokur}

\affil{Materials Science Division, Argonne National Laboratory, 9700
  South Cass, Argonne, Illinois 60439, USA}

\email{vinokur@anl.gov}

\begin{abstract}
  We investigate a one-dimensional electron liquid with two point
  scatterers of different strength.  In the presence of electron
  interactions, the nonlinear conductance is shown to depend on the
  current direction.  The resulting asymmetry of the transport
  characteristic gives rise to a ratchet effect, i.e., the
  rectification of a dc current for an applied ac voltage.  In the
  case of strong repulsive interactions, the ratchet current grows in
  a wide voltage interval with decreasing ac voltage.  In the regime
  of weak interaction the current-voltage curve exhibits oscillatory
  behavior.  Our results apply to single-band quantum wires and to
  tunneling between quantum Hall edges.
\end{abstract}

\begin{keywords}
  ratchets, Luttinger liquids, conductance, current rectification,
  bosonization, impurity scattering, Hartree approximation
\end{keywords}

\section{Introduction}

Systems with asymmetric transport characteristics are also known as
ratchets and Brownian motors since they can be used to generate dc
currents from an ac or noisy voltage.  This mechanism has important
applications in physics and biology \cite{appl}.  As specific
examples, we mention diodes and photovoltaic current rectifiers.  Here
we address the possibility of realizing such effects in Luttinger
liquids, which is of interest in connection with nanostructured
devices.  Motivated by recent experiments on quantum dots
\cite{Auslaender}, carbon nanotubes \cite{Postma}, and quantum Hall
systems \cite{Comforti}, we examine ratchet effects caused by two
unequal constrictions as point scatterers.

\section{Model}

We consider a one-dimensional spinless electron liquid at zero
temperature subject to an impurity potential $V(x)$ and a pair
interaction $W(x)$ with the Hamiltonian
\begin{eqnarray}
  H &=&  - \frac {\hbar^2}{2m} \int dx \ 
  \Psi^\dagger(x) \partial_x^2 \Psi(x)
  + \int dx  \ V(x) \Psi^\dagger(x) \Psi(x)
  \nonumber \\
  && + \frac 12 \int dx \int dx' \ 
  \Psi^\dagger(x) \Psi^\dagger(x') W(x-x') \Psi(x') \Psi(x) \ .
  \label{H}
\end{eqnarray}
The impurity potential $V(x)$ is chosen to describe two unequal point
scatterers at a distance $a$.  In the absence of interactions, the
application of a voltage $U$ leads to a current $I$ with a finite
conductance $G=I/U$.  The conductance is proportional to
$\int_{E_F-eU/2}^{E_F+eU/2} dE \ T(E)$ with the transmission
probability $T(E)$ for electrons with incident energy $E$ in the
vicinity of the Fermi energy $F_F$.  As a consequence of time reversal
symmetry, $T(E)$ does not depend on the direction of the incoming
momentum and therefore noninteracting electrons have a symmetric
transport characteristic with an odd function $I(U)$.  Therefore, the
inclusion of interactions is mandatory for the analysis of ratchet
effects.

\section{Bosonization}

For the inclusion of interactions, the bosonized representation of the
model is particularly convenient.  The bosonization technique
\cite{Voit} maps the quantum dynamics of the electron liquid onto a
path integral for a bosonic field $\vartheta$ which essentially
describes collective displacements of the electron liquid.  In terms
of this field, the particle density of the electrons reads
\begin{eqnarray}
  \rho(x) = \frac 1{\sqrt \pi} \partial_x \vartheta(x) 
  +  \frac 1{2\pi \alpha} \left( e^{i [2 k_F x +  2 \sqrt \pi \vartheta(x)]}
    + \rm{h.c.} \right) \ .
\label{rho}
\end{eqnarray}
Here, $k_F$ is the Fermi wave vector and $\alpha$ is a microscopic
cutoff length scale of the order of $k_{\rm F}^{-1}$.  The particle
current is given by $j = - \frac1{\sqrt\pi} \langle\partial_t
\vartheta \rangle$, and the charge current by $I=ej$.

In terms of the field $\vartheta$, the relevant contributions to the
action corresponding to the Hamiltonian (\ref{H}) read
\begin{eqnarray}
  S &=& \int dt dx \bigg\{ \frac{\hbar}{2 v_F}(\partial_t \vartheta)^2 -
  \frac{\hbar v_F}{2 g^2} (\partial_x \vartheta)^2 
  \nonumber \\ && 
  -  \frac 1{2\pi\alpha} V(x) 
  [e^{i [2 k_F x +  2 \sqrt \pi \vartheta(x)]} + {\rm h.c.} ]
  \nonumber \\ &&
  - \frac{W_0}{\pi^{3/2} \alpha} \partial_x \vartheta (x)
  [e^{i [2 k_F x + 2 \sqrt \pi \vartheta(x)]} + {\rm h.c.} ] \bigg\} \ .
\label{S}
\end{eqnarray}
Forward scattering by the interaction is included in the ``free''
bosonic theory via the Luttinger parameter $g:=[1+ W_0/(\pi \hbar
v_F)]^{-1/2}$ which is less than unity for repulsive interaction.
Assuming that $W(x)$ is short ranged due to screening effects, only
its weight $W_0= \int dx \ W(x)$ is effective.  The second and third
lines of Eq. (\ref{S}) describe backscattering off the impurity
potential and by the interaction, respectively.

For a single point-like scatterer, $V(x) \simeq V_0 \delta(x)$, it was
shown \cite{KF,FN,MYG} that repulsive electron interactions ($g<1$)
lead to a vanishing linear conductance.  A weak scatterer was found to
suppress the conductance according to
\begin{eqnarray}
  \Delta G \sim - V_0^2 U^{2g-2} \ ,
\label{dG}
\end{eqnarray}
where $\Delta G := G - G_0$ is conductance change due to the presence
of the scatterer.  On the other hand, for a strong scatterer, the
conductance vanishes like $G \simeq t^2 U^{2/g-2}$ with the amplitude
$t$ for hopping across the scatterer.  We subsequently focus on the
weak-barrier case, deferring the strong currugation limit to Ref.
\cite{FSV}.

\section{Hartree picture}

The suppression of the conductance in the presence of interactions can
be attributed to the backscattering of electrons from a Hartree-type
potential caused by Friedel oscillation in the vicinity of the
impurity.  In the framework of the bosonic action (\ref{S}), this
backscattering arises from the second-line contribution, and the
third-line contribution is irrelevant for the asymptotics (\ref{dG}).

To illustrate the mechanism at work, it is instructive to briefly
recall the scattering features of single electrons by a double barrier
$V(x) = V_l \delta(x+a/2) + V_r \delta (x-a/2)$.  We assume the
amplitudes $V_l$ and $V_r$ to be positive.  $a$ is the distance
between the two barriers.  Comparing the probability density
$|\psi_\rightarrow(x)|^2$ for a particle incident from $x=-\infty$
with wave vector $k>0$ to the density $|\psi_\leftarrow(x)|^2$ for a
particle incident from $x=\infty$ with wave vector $-k$, one observes
that the densities are not only different but also {\it not} related
by reflection symmetry, $|\psi_\rightarrow(x)|^2 \neq
|\psi_\leftarrow(-x)|^2$ (cf. Fig. 1).  This implies that the
contributions to the Hartree potential for both cases also will not
display this symmetry.  Therefore, in a current carrying state the
effective scattering potential will depend on the current direction
such that the net amount of scattering depends on the current
direction.  Therefore, one expects a ratchet effect from the interplay
between asymmetric scattering potentials and interactions.

\begin{figure}[htbp]
  \centering
  \epsfig{file=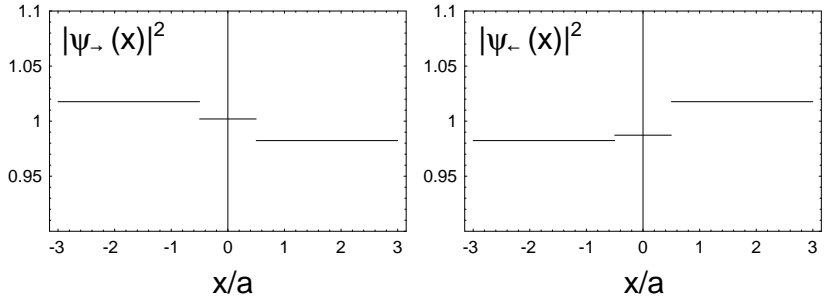,width=0.9\textwidth}
  \caption{%
    Density profiles $|\psi_{\rightarrow,\leftarrow}(x)|^2$ averaged
    over a period $\pi/k$ for a potential with $V_l < V_r$.  The
    averaged densities show drops at the positions $x=\pm a/2$ of the
    scatterers.  The amplitudes of the density drops depend on the
    direction of the incident wave.}
\end{figure}

Coming back to the electron liquid in the bosonic representation, we
find it important to include the third-line contribution in the action
(\ref{S}).  In this term, the Hartree approximation amounts to
replacing $\partial_x \vartheta$ -- which is proportional to the
slowly varying part of the density, cf. Eq. (\ref{rho}) -- by its
average value $\langle \partial_x \vartheta \rangle$.  This average is
readily calculated from the action perturbatively to second order in
$V$.  Then, the third-line contribution can be absorbed into the
second-line contribution by replacing the bare potential $V(x)$ with
the effective potential $\tilde V(x) = V(x) + \delta V(x)$ including
the correction $\delta V(x)= 2 W_0 \langle \partial_x \vartheta(x)
\rangle /\sqrt\pi$.  In analogy to the analysis \cite{KF} yielding Eq.
(\ref{dG}) for a single scatterer, one then can calculate the
corrections to conductance to second order in this effective
potential.

To estimate the strength of this effect, it is important to notice
that, to leading order, this correction is proportional to the
backscattering current off the impurities, i.e., one expects $\delta V
\propto V^2 U^{2g-1}$ corresponding to Eq. (\ref{dG}).  The
reinsertion of this correction into Eq. (\ref{dG}) suggests subleading
ratchet corrections to the conductance of order $\Delta G|_{\rm ratch}
\sim V^3 U^{4g-3}$.

Performing the systematic calculation outline above, we obtain an
asymmetric ratchet contribution to the conductance
\begin{eqnarray}
  \Delta G|_{\rm ratch} &\propto& 
  W_0 U^{4g-2} \sin(2 k_F a) V_l V_r (V_l-V_r) 
  \nonumber \\ && \times
  H_g(\frac{geUa}{\hbar v_F}) [ 1- \cos(2k_Fa)  H_g(\frac{geUa}{\hbar v_F})]
\label{res}
\end{eqnarray}
with the function
\begin{eqnarray}
  H_g(z) = \sqrt \pi \frac{\Gamma(2g)}{\Gamma(g)} 
  \frac{J_{g-1/2}(z)}{(2z)^{g-1/2}}
\end{eqnarray}
invoking the Bessel function $J$.

Equation (\ref{res}) was obtained within the Hartree approximation. It
becomes exact in a model of many bands $i$ interacting through the
coupling ${\rm interaction}_{ij}={\rm const.}\int \partial_x\phi_i
\cos(2xk_F+2\sqrt{\pi}\theta_j)$.  However, the order-of-magnitude
estimate of the current is valid in a more general case including our
one-band model (\ref{S}). This can be verified by analyzing the
expression for the current in the forth order of the perturbation
theory.

\section{Conclusions}

Equation (\ref{res}) is our main result.  Its proportionality to $W_0$
reflects the fact that the ratchet effect vanishes in the absence of
interactions.  It is valid provided the backscattering current can be
obtained within perturbation theory.  This is the case for weak
scattering, more specifically for
\begin{eqnarray}
  V/E_F \ll (eU/E_F)^{1-g}.
\label{valid}
\end{eqnarray}
Within this limit, one can distinguish two regmies.
\begin{itemize}
\item [(i)] At lower voltages $geUa \ll \hbar v_F$, $H_g (z) \approx1$
  and the effect is in agreement with the above estimate.  Additional
  oscillating factors reflect resonances due to quantum interferences
  \cite{others}.  The absolute value of the ratchet current grows with
  decreasing voltage for $g<1/2$.
\item [(ii)] In the high-voltage regime $geUa \gg \hbar v_F$, $H_g(z)
  \sim z^{-g} \cos(z-\pi g/2)$. Then $\Delta G|_{\rm ratch} \propto
  U^{3g-3}$.  In this regime, the ratchet current grows with
  decreasing votalge for $g<2/3$.
\end{itemize}
This result shows that the ratchet effect can be increasing with
decreasing voltage.  At low voltages beyond the point where the
condition (\ref{valid}) breaks down and the total current vanishes
with decreasing voltage, also the ratchet current has to vanish.
Nevertheless, it can give a sizeable contribution to the total
current.  This explains the pronounced asymmetry observed
experimentally in corrugated nanotubes \cite{Postma}.

Although the Hamiltonian (\ref{H}) does not describe quantum Hall
systems, the bosonized form (\ref{S}) captures tunneling of electron
or quasiparticles between edges \cite{Chamon}.  In a fractional
quantum Hall state with filling factor $\nu \ll 1$, the tunneling of
quasiparticles corresponds to the case $g=\nu$.  Thus, the interesting
regime with small $g$ is physically accessible.

\end{document}